\documentclass[12pt]{article}
\usepackage[latin1]{inputenc}
\usepackage[T1]{fontenc}
\usepackage{amsmath}
\usepackage{amsfonts}
\usepackage{amssymb}
\usepackage{latexsym}
\usepackage[english]{babel}
\newcommand{\be}{\begin{equation}}
\newcommand{\ee}{\end{equation}}
\newcommand{\beq}{\begin{equation}}
\newcommand{\eeq}{\end{equation}}
\newcommand{\bea}{\begin{eqnarray}}
\newcommand{\eea}{\end{eqnarray}}

\def\be{\begin{equation}}
\def\ee{\end{equation}}
\def\ba{\begin{eqnarray}}
\def\ea{\end{eqnarray}}

\begin{document}
\title{Area terms in entanglement entropy}
\author{Horacio Casini, F.D.  Mazzitelli and Eduardo Test\'e\\{\sl Centro At\'omico Bariloche,
8400-S.C. de Bariloche, R\'{\i}o Negro, Argentina}}
\date{}
\maketitle

\begin{abstract}
We discuss area terms in entanglement entropy and show that a recent formula by Rosenhaus and Smolkin is equivalent to the term involving a correlator of traces of the stress tensor in Adler-Zee formula for the renormalization of the Newton constant. We elaborate on how to fix the ambiguities in these formulas: Improving terms for the stress tensor of free fields, boundary terms in the modular Hamiltonian, and contact terms in the Euclidean correlation functions. We make computations for free fields and show how to apply these
calculations to understand some results for interacting theories which have been studied in the literature. We also discuss an application to the F-theorem. 
\end{abstract}

\section{Introduction}

The black hole entropy formula has had a large impact in theoretical physics, prompting to the discovery of the holographic nature of quantum gravity \cite{holo} and also influencing rather distant areas of research. As an example, ideas about area laws of entropy for fundamental states have been relevant in recent applications of quantum information theory to approximation methods for complex fundamental states in many body physics \cite{area1}.
    
The possibility that the statistical origin of black hole entropy can be explained as entanglement entropy (EE) across the horizon \cite{bombelli} has led to a wealth of investigations on statistical aspects of vacuum fluctuations in quantum field theory (QFT). As it is well known, the area term of entanglement entropy in QFT is divergent and cannot be renormalized without gravity. The investigation has then focused mainly in subleading universal terms. 

However, as argued in \cite{herzberg} (see also \cite{seeo}), for regions which are large with respect to the mass scales on the theory we expect an expansion of the form
\bea
S&=&\mu L^{d-2} + \textrm{shape dependent terms}\,,\\
\mu&=&\left(\frac{k_{d-2}}{\epsilon^{d-2}}+k_{d-3} \frac{m}{\epsilon^{d-3}}+...+k_0 m^{d-2}\log(m \epsilon)+k_0^\prime m^{d-2}\right)\,.\label{ff}
\eea
The area term should not depend on the shape as long as the curvatures of the boundary are much smaller than the scale $m$. In this expression $L^{d-2}$ stands for the area of the boundary of the region, $m$ is some physical mass scale of the theory, and $\epsilon$ is a short distance cutoff. The power structure of divergences can change with interactions \cite{powers} and contain non-universal coefficients, but the coefficient $k_0$ of the logarithmically corrected area term should be universal. In absence of a logarithmic contribution this term is replaced by a universal cutoff independent area term $k_0^\prime m^{d-2}$ (which is not universal in presence of a logarithmic term). For free theories the logarithmic term appears for even dimensions while it is absent for odd dimensions \cite{herzberg}. Holographic analysis show this can change with interactions \cite{powers,powers2,powers3}. 

Hence, there are some universal area terms for non conformal theories. An unambiguous definition of these terms requires a sufficiently nice regularization \cite{nice} or using the mutual information as a ``point splitting'' regularization \cite{mutu}. 

A useful way to think about the mass corrections in $\mu$ in (\ref{ff}) is that these corrections arise in the area term as we increase the size of a  region from $L\ll m^{-1}$ to $L\gg m^{-1}$. In this picture the mass dependent area terms in $\mu$ for a large planar surface in fact measure the total change of the area term between the UV and the IR limits of the theory, and depends on the details of this RG flow.  In particular, for $d=3$,  strong subadditivity combined with Lorentz invariance imply a negative $\Delta \mu=\mu_{IR}-\mu_{UV}$ for circles \cite{rg-circle}. Interestingly, a non zero running of $\mu$ entails a non zero running of the constant term of the entropy of circles $c_0$. This later acts as the monotonic quantity of the F-theorem \cite{f-theo}.   

The idea of a running of the area term with the scale of the region together with the black hole entropy formula suggests that even if we do not understand how gravity regularizes the UV divergences and makes the entropy finite and equal to $A/4G$, a universal low energy contribution to the black hole entropy should originate in QFT and be calculable as entanglement entropy \cite{otro,otro1} (see also \cite{oot}). More precisely, the universal piece in the renormalization of the area term in EE should be the same as the one in the renormalization of $(4G)^{-1}$.

 The renormalization of the Newton constant due to integration of massive fields can be computed in the context of QFT in curved spaces, using for instance heat kernel techniques \cite{birrell}.  Alternatively, it is given by the so called Adler-Zee (AZ) formula \cite{adler,zee} in terms of flat 
 space correlators of the trace of the stress tensor $\Theta(x)=T_{\mu}^\mu(x)$,
\begin{equation}
\Delta((4 G)^{-1})=-\frac{\pi}{d(d-1)(d-2)}\int d^{d}x\,\,x^2 \,\langle 0| \Theta(0)\Theta(x)| 0\rangle+\frac{4 \pi}{d-2}\langle {\cal O}\rangle\,.\label{AZ}
\end{equation}
Here ${\cal O}=\delta \Theta/\delta R$ is the variation of the stress tensor trace with respect to the curvature for a conformally flat metric \cite{adler,hirohumi}, and it is relevant when there are non trivial couplings of the fields with the curvature. 
Everything is written in Euclidean metric. This formula follows by identifying the renormalization of the coefficient of the scalar curvature in the gravitational effective action once the fields have been integrated out.

However, as it stands, the idea that the universal part of the area terms in the entropy of quantum fields is given by (\ref{AZ}) does not quite work since the universal part of AZ formula depends on the particular way the fields are coupled to gravity, while this is not the case for $\mu$ for the QFT in flat space, which does not know about gravity. 

Important progress in understanding the area term in EE was recently produced by Rosenhaus and Smolkin \cite{rs,rs1,rs2,rs3}. They obtain a formula based on the first law of EE \cite{first-law,takaw}
\be
\mu=\frac{1}{d-2}\int d^dx\, \langle \Theta(x) K \rangle\,,\label{RS}
\ee
where $K=-2\pi \int d^{d-1}x\, x^1 T_{00}$ is the modular Hamiltonian (the rotation operator in Euclidean formulation, that is the analytic continuation of the real time boost operator) of the half plane\footnote{An expression of the renormalization of the entanglement entropy in terms of expectation values of the trace $\theta$ in a conical manifold was worked out in \cite{baner2}}. As described below, Rosenhaus-Smolkin (RS) formula is easily shown to be equivalent to the first term in AZ formula (\ref{AZ})
\begin{equation}
\mu=-\frac{\pi}{d(d-1)(d-2)}\int d^{d}x\,\,x^2 \,\langle 0| \Theta(0)\Theta(x)| 0\rangle \label{fgf}
\end{equation}
 using spectral decomposition for the stress tensor correlators. However, both of these formulas have to be treated with some care, and in the subtleties of the precise definition, differences may appear between renormalizations of $(4G)^{-1}$ and universal area terms in QFT. 

In this paper we focus in trying to understand more precisely the universal terms in $\mu$ in QFT, and the subtleties in the evaluation of the first term of (\ref{AZ}) or equivalently (\ref{RS}). Our guiding principle is that the result must be uniquely defined by the QFT in flat space, that is, by the operator content and the correlation functions. For example, for free theories there should not be any dependence on the particular choice of stress tensor, or equivalently, on couplings of the field with the curvature when the theory is extended to curved space. There should neither be ambiguities in the definition of the modular Hamiltonian operator due to boundary terms (see the recent discussions \cite{modu-ambig,modu2,malda1,herzog}) or ambiguities coming from contact terms in the expression (\ref{AZ}) at the coincident point of the operators. These contact terms are distributions with support at the contact point in the Euclidean correlators, and are not determined by analytic continuation of the real time Wightman functions of the theory \cite{OS}. Hence they should not play any role in the universal part of $\mu$. Contact terms and improvement terms however play an important role in the precise calculation of the renormalization of the Newton constant. We mainly do calculations for free fields but we were able to understand the area terms for some interacting theories that have been worked out in the literature by exploiting free field results: The $O(N)$ scalar model in the $\epsilon$ expansion discussed in \cite{on} and the absence of logarithmically corrected area term in $d=4$ for a scalar mass term computed holographically \cite{powers,powers2}. 

The plan of the paper is as follows. In the next section we show how the universal area terms can be defined using mutual information. We also compute these terms for free fields using the mutual information between two half planes. In section 3 we make a short derivation of the RS result and show the formal equivalence between their formula and AZ. In section 4 we discuss the relations between the formula for area terms to c-theorems in $d=2$ and $d=3$. In particular we argue that an area term renormalization with negative sign in $d=3$ drags the running of the c-function to be strictly positive, hence establishing a property of ``isolation of fixed points'' analogous to the one in Zamolodchikov's theorem for $d=2$. In section 5 we discuss free fermion and scalar fields, and study several subtleties, like the definition of the modular Hamiltonian and the inclusion of improvement terms. In section 6 we review the calculation of the renormalization of the Newton constant
 for the free scalar and discuss the role of couplings with the curvature and the second term in AZ formula, as well as the importance of contact terms in the first term. We apply these ideas to interacting examples in section 7. Finally, we present our conclusions.

\section{Area terms in mutual information}
\label{mutual-ap}

Mutual information $I(A,B)=S(A)+S(B)-S(A\cup B)$ for non intersecting sets $A$ and $B$ is a universal quantity, depending only on the continuous QFT, that is, it is independent of regularization. It can also be used as a particular geometric regularization of entropy, analogous to framing regularization for Wilson loops or point splitting regularization for expectation values of point operators, with 
the definition \cite{mutu}
\be
S(A,\epsilon)\sim \frac{1}{2} I(A_{\epsilon/2},A^c_{\epsilon/2})\label{gg}\,,
\ee
where $\epsilon$ is a small physical distance between the two regions, $A_{\epsilon/2}$, which is $A$ contracted a distance $\epsilon/2$ from the boundary, and $A^c_{\epsilon/2}$, which is the complement $A^c$ of $A$, contracted $\epsilon/2$ from the boundary. 

 The mutual information $I(A_{\epsilon/2},A^c_{\epsilon/2})$ for two parallel planar entangling surfaces separated by a distance $\epsilon$ has an expansion similar to (\ref{ff}), but where all terms are now universal.  The calculation of the divergent universal terms in the expansion is  difficult since it is necessary to compute the entropy for the union of two half spaces in (\ref{gg}) or equivalently the entropy for a long thin strip between the two planes. We are doing this calculation for free fields below. However, the $\epsilon$ independent term (or the logarithmic term in $\epsilon$, depending on space-time dimensions) should not retain information on the finite width strip, and hence be assimilated to the universal part in the half space entanglement entropy for a sufficiently geometric regularization \cite{nice}.    

The discussion of mutual information will also serve to understand boundary terms in the modular Hamiltonian for the half space later.
 
For free fields we can make the calculation by dimensional reduction. We take two half spaces, say for $ x^1< -\epsilon/2$ and for $x^1>\epsilon/2$, separated by a distance $\epsilon$. For a scalar field for example, the field can be Fourier transformed in the directions parallel to the plane, and we have the Hamiltonian
\bea
H&=&\frac{1}{2}\int d^{d-1}x\, \left(\pi^2(x)+(\nabla \phi(x))^2\right)+m^2 \phi^2(x))\nonumber\\
&=&\left(\frac{L}{2\pi}\right)^{d-2}\int d^{d-2}p_{\parallel} \,\frac{1}{2} \,\int dx^1\,\left(\pi_{p_{\parallel}}^2(x)+(\partial_{x^1} \phi_{p_{\parallel}}(x))^2+(m^2+p_{\parallel}^2) \phi_{p_{\parallel}}(x)\right)\,.
\eea

Mutual information for the vacuum state can then be decomposed in the different momentum of the direction parallel to the planes, and the result is given by an integral of the two dimensional mutual informations in the $1+1$ spacetime of coordinates $x^0,x^1$ for the tower of two dimensional fields with effective masses $M^2=m^2+p_{\parallel}^2$. These are in turn given in terms of the one dimensional entropic c-functions. Details are in \cite{uno,dos}. We have
\bea
I(\epsilon,m)=\kappa L^{d-2}\, m^{d-2} \int_{m \epsilon}^\infty dx\, \frac{1}{x^{d-1}}\int_0^\infty dq\, q^{d-3} C(\sqrt{q^2+x^2} )\nonumber\\
=\kappa L^{d-2}\, m^{d-2} \int_{m \epsilon}^\infty dx\, \frac{1}{x^{d-1}}\int_{x}^\infty dy\, y \,(y^2-x^2)^{\frac{d-4}{2}}\, C(y)\,.\label{dise}
\eea
Here $C(M \epsilon)=\epsilon dS(\epsilon,M)/d\epsilon$ is the dimensionless c-function of a two dimensional field of mass $M$ \cite{uno,tres}, and 
\be
\kappa=\frac{d-2}{2^{d-2}\pi^{\frac{d-2}{2}}\Gamma[d/2]}\,.
\ee 

According to (\ref{dise}) we need to expand the last integral in powers of $x$ to get an expansion of $I(\epsilon,m)$ in powers of $\epsilon$. For even dimensions the coefficient function inside the integral is a polynomial, and we can expand in $x$ directly
\be
\int_{x}^\infty dy\, y \,(y^2-x^2)^{\frac{d-4}{2}}\, C(y)=\sum_{i=0}^{d/2-2} (-1)^i \left(
\begin{array}{c}
 \frac{d}{2}-2 \\
 i 
\end{array}
\right)C^{(d-3-2i)} x^{2 i}+(-1)^{\frac{d}{2}-1} \frac{C(0) x^{d-2}}{d-2}+ {\cal O}(x^{d-1})\,,
\ee
where $C^{(n)}$ is the $n^{\textrm{th}}$ momentum of the c-function,
\be
C^{(n)}=\int_0^\infty dy\, y^n \, C(y)\,.
\ee
The terms in the sum come from expanding the polynomial in the integrand and taking $x\rightarrow 0$ in the lower limit of the integral. The last term proportional to $x^{d-2}$ comes from deriving $d-2$ times with respect to $x$, and evaluating at $x=0$.

Plugging this back into (\ref{dise}) we obtain a series expansion for $I(\epsilon,m)$ for small $m\epsilon$,
\be
I(\epsilon,m)=L^{d-2}\left(\sum_{i=0}^{d/2-2} \frac{(-1)^{i} k}{d-2 i-2} \left(
\begin{array}{c}
 \frac{d}{2}-2 \\
 i 
\end{array}
\right)C^{(d-3-2i)} \frac{m^{2 i}}{\epsilon^{d-2-2 i}}+(-1)^{\frac{d}{2}}\frac{k C(0) m^{d-2}\log(m \epsilon)}{d-2}\right)\,.\label{tr}
\ee
This equation gives an expansion that contains all the divergent terms as $\epsilon$ goes to zero. Notice that only even powers of $\epsilon$ appear. The coefficients depend on the moments of the c-function, which are slightly different for a scalar and a Dirac fermion. They can be computed numerically with high precision using the expression of $C(x)$ in terms of solutions of non-linear differential equations \cite{uno,tres}. In contrast, the logarithmic term is proportional to the value of the c-function at the origin, which is $1/3$ both for Dirac fermions and real scalars. Because of this dependence on the function $C(x)$, the coefficients of the non-universal divergent terms in the entropy will not be the same as the ones computed with other regularizations, e.g. heat kernel \cite{herzberg}. However, the coefficients of the mutual information are physical for all powers.  Dividing by two the result of the mutual information we obtain the logarithmic term for a scalar field in even dimensions 
\be
\mu^S_{\textrm{univ}}=(-1)^{\frac{d}{2}}\frac{1}{2^{d-1} 3 \pi^{\frac{d-2}{2}}\Gamma[d/2]}m^{d-2}\log(m \epsilon)\,.\label{evend}
\ee
This is exactly the result for $\mu_S$ obtained in \cite{herzberg} by the heat kernel method. 

The fermion has half the coefficient of the scalar per spinor component, because it inherits this from the relation of $C$ for the massless $d=2$ case, where we have $C(0)=1/3$ both for Dirac and real scalar fields (one half the scalar result for each fermion component). Hence we get
\be
\mu^F_{\textrm{univ}}=\frac{d_\Psi}{2} \mu^S_{\textrm{univ}}\,,\label{ferm}
\ee
where $d_\Psi$ is the number of spinor components. This also coincides with heat kernel calculations \cite{powers2}.

For a scalar field, due to the cuspy behavior of the c-function at the origin \cite{dos}
\begin{equation}
C(x)\sim \frac{1}{3}+\frac{1}{2}\frac{1}{\log(x)}+...\,,
\end{equation}
 there is an additional  
 \be
 \frac{(-1)^{d/2} m^{d-2}}{2(d-2)}\log (-\log(m \epsilon))
 \ee
  term in $\mu$ for even dimensions.  However, this does not affect the universality of the logarithmic term for even $d$. 

For the case of odd dimensions the calculation of the power terms also proceeds by expanding (\ref{dise}), with the same result as in the power terms in (\ref{tr}) where the binomials for $d$ odd are now
\be
\left(
\begin{array}{c}
 \frac{d}{2}-2 \\
 i 
\end{array}
\right)=\frac{(d/2-2)(d/2-2-1)...(d/2-2-(i-1))}{i!}\,.
\ee   

The constant term comes from extracting the power terms and evaluating the integrals for $m\epsilon=0$. This is
\bea
&&k L^{d-2}\, m^{d-2} \int_0^\infty dy\,\, C(y) \int_0^\infty dx\, \left(\theta(y-x) \frac{y \,(y^2-x^2)^{\frac{d-4}{2}}}{x^{d-1}}\right. 
\nonumber\\ &&\hspace{6cm}\left.-\sum_{i=0}^{d/2-3/2} (-1)^i \left(
\begin{array}{c}
 \frac{d}{2}-2 \\
 i 
\end{array}
\right)y^{(d-3-2i)} \frac{1}{x^{d-1-2 i}}\right)\label{doce}\,,
\eea
where $\theta(x)$ is the step function. 
Now, the second integral in $x$ gives exactly zero for any non zero $y$. However, the whole integral is non zero since  
\bea
&&\int_0^\infty dx\,\,\int_0^{y_0} dy\,\,\left(\theta(y-x) \frac{y \,(y^2-x^2)^{\frac{d-4}{2}}}{x^{d-1}} -\sum_{i=0}^{d/2-3/2} (-1)^i \left(
\begin{array}{c}
 \frac{d}{2}-2 \\
 i 
\end{array}
\right)y^{(d-3-2i)} \frac{1}{x^{d-1-2 i}}\right)\nonumber \\
&&=(-1)^{\frac{d-1}{2}}\frac{\pi}{2(d-2)}\,,
\eea
independently of the value of $y_0$. Hence, the $x$ integral in (\ref{doce}) is proportional to a delta function $\delta(y)$. The constant term in odd dimensions will depend on the value of $C(0)$ alone, as is the case of the logarithmic term in even dimensions. With this we have the full expansion for odd $d$ 
\be
I(L,m)=L^{d-2}\left(\sum_{i=0}^{d/2-3/2} \frac{(-1)^{i} k}{d-2 i-2} \left(
\begin{array}{c}
 \frac{d}{2}-2 \\
 i 
\end{array}
\right)C^{(d-3-2i)} \frac{m^{2 i}}{\epsilon^{d-2-2 i}}+(-1)^{\frac{d-1}{2}}\frac{k \pi C(0) m^{d-2} }{2 (d-2)}\right)\,.\label{tr1}
\ee 

Using $C(0)=1/3$ for a scalar, the coefficient of the constant term in the entropy is 
\be
\mu^S_{\textrm{univ}}=(-1)^{\frac{d-1}{2}}\frac{\pi}{2^{d} 3 \pi^{\frac{d-2}{2}}\Gamma[d/2]} m^{d-2}\,,\label{oddd}
\ee
and we again have the relation (\ref{ferm}) for fermions. This result coincides with the ones previously obtained in the literature \cite{herzberg,powers2}.

There is no  $\log(\log(-m\epsilon))$ term for the scalar in odd dimensions. This is good, otherwise the presence of this term in odd dimensions would have spoiled the universality of the constant term.

\section{Formulas for the area term}

The variation of the entanglement entropy under a small change of the state is given by
\be
\delta S=\textrm{tr}(\delta \rho K)\label{44}\,,
\ee
with $\delta S=S(\rho)-S(\rho_0)$, $\delta \rho=\rho-\rho_0$, and where $K$ is the modular Hamiltonian defined by $\rho_0\sim e^{-K}$ \cite{first-law}. The density matrix can be expressed as a path integral in Euclidean space with boundary values on both sides of a cut at $t=0$ on the location of the region $V$ (see for example \cite{cala}). The modular Hamiltonian $K$ is an operator insertion in (\ref{44}) precisely at this cut at $t=0$, $\vec{x}\in V$. The vacuum state corresponds to unperturbed Euclidean space in the path integral, while boundary conditions or external sources can be used in the path integral to choose a different state. Rosenhaus and Smolkin use this representation to obtain a formula for
infinitesimal variations of the entanglement entropy under changes of relevant operators in the action. If the action contains a term $ \beta \int d^{d}x\, O(x)$  they get \cite{rs,rs1}
\be
\frac{\partial S}{\partial \beta}=-\int dx^d\,\langle O(x) K\rangle_\beta \,. \label{rp}
\ee 
Hence the variation of the entropy is given by the integral of the expectation value of the operator over Euclidean space with the modular Hamiltonian inserted at the cut. It is important to keep in mind that the use of the first law of EE (\ref{44}) is for small enough variations of the state, and this can only be achieved with a change in the action if a cutoff is in place, that is, if we make $\delta \beta$ going to zero first and then take the limit $\epsilon \rightarrow 0$. While the opposite order of limits is the correct one for continuum QFT, this should not be a problem for the calculation of the universal cutoff independent terms.  

In order to obtain the universal area term we use the fact that the entropy for a planar entangling surface has the form $S=L^{d-2} \mu$. Then we have for the variation of $L$
\be
L \frac{dS}{dL}=(d-2) S\,.\label{mira}
\ee
Now a variation of the size of the region can be obtained by changing $x\rightarrow \lambda x$ in the path integral involved in the density matrix in eq.(\ref{44}), and keeping all mass parameters and the coordinate size of the region $L$ fixed. This step is equivalent to pull down from the action ${\cal A}=\int d^dx\, {\cal{L}} $ the quantity
\be
\int d^dx\,\left(\sum_i \partial_\mu \phi^i \frac{\partial {\cal L}}{\partial (\partial_\mu \phi^i)}-d \,{\cal L}\right)=\int d^dx\,g_{\mu\nu}\left(\sum_i \partial_\nu \phi^i \frac{\partial {\cal L}}{\partial (\partial_\mu \phi^i)}-g^{\mu\nu} \,{\cal L}\right)\,.\label{47}
\ee
For free fields the quantity within brackets is the trace of the canonical stress tensor. For interacting fields having a unique stress tensor a variation of the effective action should give $\Theta=g_{\mu\nu} T^{\mu\nu}$. 
Then we get
\be
\left.\lambda\frac{\partial S}{\partial \lambda}\right|_{\lambda=1}=\int dx^d\,\langle \Theta(x) K\rangle \,. 
\ee
This equation holds only for the universal terms since we have left unspecified how the cutoff changes with the transformation. 
Since $L \frac{dS}{dL}=\lambda\frac{\partial S}{\partial \lambda}\vert_{\lambda=1}$ we have from (\ref{mira})
\be
 S=\frac{1}{(d-2)}\int dx^d\,\langle \Theta(x) K\rangle =-\frac{2\pi}{d-2}\int d^dx\, \int_{y^1>0} d^{d-2}y\,y^1 \langle \Theta(x)T_{00}(y)\rangle\,.
\ee
The coefficient of the area term is then given by
\be
\mu=-\frac{2\pi}{d-2}\int d^dx\, \int_{y^1>0} dy^1\,y^1 \langle \Theta(x)T_{00}(y)\rangle\,.\label{mu1}
\ee
This is RS formula which the authors obtain from (\ref{rp}) using a Ward identity.
The universal piece has to be extracted regularizing this divergent expression. 

An important simplification of this formula was found in \cite{rs1} by using the spectral representation of the stress tensor correlation functions \cite{cappelli},
\be
\langle T_{\alpha\beta}(x) T_{\rho\sigma}(0)\rangle=\frac{A_d}{(d-1)^2}\left(\int ds\, c^{(0)}(s) \Pi^{(0)}_{\alpha \beta,\rho\sigma}(\partial)\,G_0(x,s)+\int ds\, c^{(2)}(s) \Pi^{(2)}_{\alpha \beta,\rho\sigma}(\partial)\,G_0(x,s)\right)\,,
\ee
with
\be
A_d=\frac{\pi^{d/2}}{(d+1)\Gamma(d/2)2^{d-2}}\,,
\ee
and where $\Pi^{(0)}_{\alpha \beta,\rho\sigma}$ and $\Pi^{(2)}_{\alpha \beta,\rho\sigma}$ are fourth-order polynomial tensors in derivatives $\partial$, $c^{(0)}(s)$ and $c^{(2)}(s)$ are the spin zero and two spectral functions, and $G_0(x,s)$ is the free scalar Green function of mass $s$,
\be
G_0(x,s)=\frac{1}{2\pi}\left(\frac{s}{2\pi |x|}\right)^{(d-2)/2}K_{(d-2)/2}(s |x|)\,.
\ee
In particular $\Pi^{(2)}_{\alpha \beta,\rho\sigma}$ is traceless and hence $c^{(2)}(s)$ does not enter into the formula (\ref{mu1}). The tensor $\Pi^{(0)}_{\alpha \beta,\rho\sigma}$ is
\be
\Pi^{(0)}_{\alpha \beta,\rho\sigma}=\frac{1}{\Gamma(d)}(\partial_\alpha \partial_\beta-\delta_{\alpha\beta}\partial^2)(\partial_\rho \partial_\sigma-\delta_{\rho\sigma}\partial^2)\,.
\ee
 Eq. (\ref{mu1})  becomes simply\footnote{Note that for theories where the stress tensor operator is non unique, in order to obtain this expression from (\ref{mu1}) $\Theta(x)$ and $T_{00}(x)$ must involve the same (symmetric and conserved) stress tensor operator for the spectral representation to be valid.} \cite{rs1}
\be
\mu=-\frac{2\pi A_d}{(d-1)(d-2)\Gamma(d)}\int_0^\infty ds\,\, c^{(0)}(s)\,.\label{peri}
\ee

Interestingly, by a simple application of the spectral representation we can show that RS formula is equivalent to the term involving correlators of $\Theta(x)$ in the AZ formula for the renormalization of the Newton constant.
We have
\be
\langle\Theta(0)\Theta(x)\rangle=\frac{A_d}{\Gamma(d)}\int_0^\infty ds \, c^{(0)}(s)\,s^4 G(x,s)\,.
\ee
Then
\begin{eqnarray}
\mu&=&-\frac{\pi}{d(d-1)(d-2)}\int d^{d}x\,\,x^2 \,\langle 0| \Theta(0)\Theta(x)| 0\rangle\nonumber\\
&=&-\frac{\pi A_d}{d(d-1)(d-2)\Gamma(d)}\int d^{d}x\,\,x^2 \,ds \, c^{(0)}(s)\,s^4 G(x,s)\nonumber\\
&=&-\frac{2\pi   A_d}{(d-1)(d-2)\Gamma(d)}\int_0^\infty ds \, c^{(0)}(s)\,,\label{fjkll}
\end{eqnarray}
coinciding with (\ref{peri}).
In the last step we have used
\be
\int d^dx \, G_0(x,s) \,x^2=\frac{2 d }{s^4}\,.
\ee

It is interesting to note that the expression (\ref{fgf}) gives a formally negative term in any dimensions, and the same sign for the renormalization of the area term follows from heuristic application of strong subadditivity \cite{rg-circle}. Divergences generally spoil this naive positivity argument in $d>3$.  

\section{Relations to c-theorems in $d=2$ and $d=3$}
Let us first discuss the connection of the formula for the area term with the c-theorem in $d=2$. 
The formula (\ref{fgf}) cannot be directly applied to two dimensions due to the $(d-2)$ factor at the denominator. However, we can use the following trick to dimensionally continue this formula down to $d=2$. We can extract from the integral of the correlation functions a mass parameter to make it dimensionless
\be
\mu=-\frac{\pi\,\,m^{d-2}}{d(d-1)(d-2)}  \int d^{d}x\,\,x^2 \,m^{-(d-2)} \langle 0| \Theta(0)\Theta(x)| 0\rangle\,.
\ee
In the limit $d\rightarrow 2$ this gives a universal logarithmic term
\be
\mu=-\frac{\pi}{2}  \left(\int d^{2}x\,\,x^2 \langle 0| \Theta(0)\Theta(x)| 0\rangle\right) \log(m)\,.\label{pol}
\ee
The integral within brackets is dimensionless in $d=2$ and is given in terms of the change in the Virasoro central charge $C_V$ between the ultraviolet and infrared fixed points (see for example \cite{cappelli}) 
\begin{equation}
\Delta C_V=C_V^{UV}-C_V^{IR}=3 \pi\int d^{2}x\,\,x^2 \,\langle 0| \Theta(0)\Theta(x)| 0\rangle\,.\label{iot}
\end{equation}
This quantity is always positive due to the c-theorem, which in (\ref{iot}) is a manifestation of the reflection positivity of the correlator together with the convergence of the integral. 

Hence we have for (\ref{pol})
\be
\mu=-\frac{1}{6} (C_V^{UV}-C_V^{IR}) \log(m \epsilon)\,,\label{hgf}
\ee
where we have inserted a distance cutoff $\epsilon$ to compensate for the dimensions. 

This is indeed what we expect for the renormalization of the ``area term'' in $d=2$ \cite{cala}. For a small interval of size $R$ in $d=2$ the entropy is
\be
S_{UV}=\frac{C_V^{UV}}{3}\log(R/\epsilon)+k_0\,,
\ee
with $k_0$ a non universal constant.
For a large interval compared to all mass scales in the theory we should have
\be
S_{IR}=\frac{C_V^{IR}}{3}\log(R/\epsilon)+k_0^\prime- \frac{C_V^{UV}-C_V^{IR}}{3}\log(m \epsilon)\,.\label{oro}
\ee
The coefficient of the last term in this equation is fixed by the requirement that the $\log(\epsilon)$ term in both $S_{UV}$ and $S_{IR}$ must be the same. This is due to the fact that this term is generated by ultraviolet entanglement around the boundaries of the interval, and must be independent of the size (see also \cite{baner}). The physical mass scale $m$ can be chosen at will. 

The last term in (\ref{oro}) should be compared with (\ref{hgf}). In (\ref{hgf}) we have $-(C_{UV}-C_{IR})/6$ while the coefficient for a large  interval is $-(C_{UV}-C_{IR})/3$ because formula (\ref{pol}) is for the half space, which has only one boundary, while an interval has two boundaries (a double ``area'').  

Therefore we conclude that (\ref{fgf}) does indeed give the renormalization of the area term in $d=2$ for any theory.

\bigskip

In dimension $d=3$ the c-theorem (F-theorem) states that the constant term $c_0$ in the entropy of a circle decreases from UV to IR fixed points, where for a conformal point the entropy of a circle writes
\be
S(R)=R \left(\frac{k_1}{\epsilon}+k_0\right)-c_0\,.
\ee
No logarithmic term is present for the entropy of spheres in odd dimensions \cite{my}. Outside the fixed points, due to strong subadditivity and Lorentz invariance, the circle entropy satisfies \cite{rg-circle}
\be
S^{\prime\prime}(R)<0\,.
\ee
The running of the constant and area terms are given by \cite{rg-circle} 
\bea
c_0^{UV}-c_0^{IR}&=&-\int_0^\infty dR\, R\,S^{\prime\prime}(R)\ge 0\,,  \label{fb}\\
\mu=k_0^{IR}-k_0^{UV}&=&\int_0^\infty dR\, S^{\prime\prime}(R)\le 0\,.\label{fa}
\eea
Then for $d=3$ the renormalization of the area term has negative sign. 

Interestingly, this result coming from strong subadditivity is in accordance with the sign implied by reflection positivity in the formula
\be
\mu=-\frac{\pi}{6}\int d^{3}x\,\,x^2 \,\langle 0| \Theta(0)\Theta(x)| 0\rangle\,.\label{69}
\ee
In $d=3$ there can be some theories where the integral in the right hand side can be divergent, and hence the formal positivity of the correlator does not imply a negative universal term. However, the universal term in $\mu$ must be negative because of the c-theorem. We will find that this is the case for free scalars where the integral diverges because we are forced to choose the canonical stress tensor, but where the negative sign is preserved for the universal term.   

For free fermions the integral is convergent and we expect the same for most non-free fixed points and runnings. For example, if we perturb the UV CFT with a relevant scalar operator of dimension $1/2<\Delta<3$ we get $\langle 0| \Theta(0)\Theta(x)| 0\rangle\sim |x|^{-2 \Delta}$ and this gives a convergent integral in (\ref{69}) for $\Delta<5/2$. At the infrared, perturbing with an irrelevant operator of dimensions $\Delta>3$ we get always a convergent integral for large $x$. Then in principle there is no problem with divergences at the infrared. 

For the case of a perturbation with $3>\Delta> 5/2$ at the UV probably the area term does have a divergent total running and there is no universal finite area term. This is what happens holographically. In the holographic case the bulk metric is given generically by
\be
ds^2=\frac{\tilde{L}^2}{z^2}\left(-dt^2+d\vec{x}^2+\frac{dz^2}{f(z)}\right)\,,
\ee  
where $\tilde{L}$ is the asymptotic AdS radius, and $f(z)$ describes the behavior of the theory with scales. The null energy condition implies $f^\prime(z)>0$. We can set $\lim_{z\rightarrow 0}f(z)=1$ and $\lim_{z\rightarrow \infty} f(z)=f(\infty)=\textrm{constant}$. The bulk minimal surface ending at a $d-2$ dimensional plane at the $d$-dimensional boundary goes straight to the bulk direction. The entropy is then proportional to this minimal area \cite{rt}
\be
S= \frac{L^{d-2}}{4 G_N^{d+1}}\int dz \frac{\tilde{L}^{d-1}}{f^{1/2}(z) z^{d-1}}\,.\label{va}
\ee
Near $z=0$ we have an expansion $f(z)=1+(\tilde{\mu} z)^{2(d-\Delta)}$ for a CFT perturbed by an operator of dimension $\Delta$, and where $\tilde{\mu}$ is some mass scale. This gives a variation of the area (\ref{va}) at the UV which diverges for $\Delta>(d+2)/2$ and gives a logarithmic term for $\Delta=(d+2)/2$ \cite{powers,rs3}. This is $5/2$ for $d=3$.
   
Note that a divergent area renormalization does not imply a divergent $c_0$ renormalization since the variation of $c_0$ is suppressed precisely at the UV by an additional power of $R$ in (\ref{fb}) as compared to (\ref{fa}). 

The renormalization of the area term and the constant term both depend on the second derivative $S^{\prime\prime}(R)$, eqs. (\ref{fb}) and (\ref{fa}). These formulas however do not constraint much the relation between $\Delta c_0$ and $\mu$ except for the following. If $\mu\neq 0$ then necessarily $S^{\prime\prime}\neq 0$ (and hence $S^{\prime\prime}< 0$) for some $R$. This implies that $\Delta c_0>0$ is not zero. Therefore a non zero renormalization of the area term implies a non zero renormalization of the constant term $c_0$ in the circle entropy. 

In the light of this commentary, the interesting point about formula (\ref{69}) for the area term is that it gives $\mu$ as an integral over a correlator of the operator $\Theta$ with itself, which cannot be zero unless the operator itself is zero. Hence, if the theory is outside a fixed point, and $\Theta(x)\neq 0$, it must be necessarily that $\langle 0| \Theta(0)\Theta(x)| 0\rangle>0$ for some $x$. This then implies that $\mu<0$  and that $\Delta c_0>0$. Hence, we have a result analogous to the one that follows from (\ref{iot}) for the Zamolodchikov theorem in $d=2$: In the running from a UV to an IR fixed point the central charge not only decreases but it cannot remain constant. 

Of course, for $d=3$ (as opposed to the case $d=2$) this result is complicated by the possibility of divergences in the expression for $\mu$ in some theories. In principle these can lead to a universal part of $\mu=0$. However, according to our previous discussion, holographic entanglement entropy suggest that in the case where (\ref{fgf}) is divergent, the total area running is also divergent, leading still to a positive $\Delta c_0>0$.     

\section{Free fields: Modular Hamiltonians and improvement term}

In this section we study  free massive fermion and scalar fields, and focus our attention on solving possible ambiguities in the area term formulas in order to obtain the correct universal terms given by mutual information regularization. 

\subsection{Modular Hamiltonian}

RS formula contains the modular Hamiltonian $K$ of the Rindler wedge and the derivation uses the first law $\Delta \langle K\rangle=\Delta S$ for small deviations. Recently there have been discussions in the literature about possible boundary terms in the modular Hamiltonian involved in the first law of EE \cite{modu-ambig,modu2,malda1,herzog}. Regarding this point, let us first note that the (real time) modular Hamiltonian is involved in the following one parameter group of unitary transformations, the modular flow,
\be
U(\tau)=e^{-i K \tau}\,.
\ee
This may be thought as a time evolution operator for the state $\rho_0=e^{-K}$, which is ``thermal'' with respect to the Hamiltonian $K$. Hence the correlation functions of operators localized inside the region have to satisfy the $KMS$ condition of periodicity in imaginary time \cite{haag}
\be
\langle {\cal O}_1 U(i){\cal O}_2 U(-i)\rangle=\langle {\cal O}_2  {\cal O}_1\rangle\,.
\ee  
This defines the modular flow and implies that any ambiguity in $K$ has to be given by additions of operators which commute with the operators inside the region, in such a way to keep the modular flow intact. These possible additional terms have then to be localized at the boundary. 

Another way to see this is using the relative entropy
\be
S(\rho|\rho_0)=\Delta \langle K\rangle-\Delta S\,.
\ee
This is a universal quantity \cite{ros} for any region and state.
Now, if we take a unitary operator $U_V$ localized inside the region $V$ and as $\rho_V$ the reduced state corresponding to $U|0\rangle$, we have $\Delta S=0$ and 
\be
S(\rho_V|\rho_V^0)=\Delta \langle K\rangle\,.
\ee  
Hence $\Delta \langle K\rangle$ is uniquely defined for any excitation generated by a localized unitary operator. 

The question is whether there are ambiguities in the form of boundary terms for $K$. According to the universality of relative entropy this is equivalent to ask if there are ambiguities in $\Delta S$ in the continuum limit. We do not have for $\Delta S$ the monotonicity properties which ensure the universality of relative entropy, and ultimately this is an open question which deserves further study.\footnote{Note that in any case possible ambiguities in $\Delta S$ are highly constrained in the continuum to be expectation values of operators localized at the boundary, while the entropy is a non linear functional of the density matrix.} However, heuristically, ultraviolet terms should cancel in $\Delta S$ and we should get a universal quantity. $\Delta S$ could be defined using mutual information regularization for $S$, giving a universal prescription for it, and hence for $K$.   

Let us take the case of a free massless scalar in Rindler space. The modular flow is given by boost operators. While the boost generator is uniquely defined this is not so for the boost generator restricted to one half of the space. 
 The stress tensor for the scalar can be chosen from the family of conserved $d$-dimensional tensors
\begin{equation}
T_{\mu\nu}=\partial_\mu\phi \partial_\nu \phi-\frac{1}{2} g_{\mu\nu}\left(\partial_\alpha \phi\partial^\alpha \phi\right)-\xi \left(\partial_\mu\partial_\nu-g_{\mu\nu}\partial_\alpha\partial^\alpha\right)\phi^2\,,
\end{equation}
where $\xi$ is a free parameter describing the coupling of the field with the curvature when the metric is deformed. The case
with
\begin{equation}
\xi=\xi^c\equiv\frac{(d-2)}{4 (d-1)}\label{confor}
\end{equation}
gives the conformal tensor with $T_\mu^{\mu}=0$.
In flat space, however, the theory is always the same and $\xi$ only gives an arbitrary choice of stress tensor. All these tensors give the same boost generator but 
\be
K_\xi=2\pi\int_{x^1>0} d^{d-1}x\,x^1\,T_{00}(x,\xi)=K_0-2 \pi \xi \int_{x^1=0}d^{d-2}x\,\phi^2(x)\,, \label{case}
\ee  
where $K_0$ is the ``half boost'' generator for the canonical (or minimally coupled) choice $\xi=0$.

The discussion of the calculation of mutual information in section 2 by dimensional reduction shows that the modular Hamiltonian for half space for the free scalar is given by the sum of the modular Hamiltonians for the region $x^1>0$ of massive fields in dimension $d=2$. Therefore, the right choice in dimension $d$ cannot depend on the dimension, as would be the case with (\ref{case}) for the conformal choice (\ref{confor}). In $d=2$ the canonical stress tensor coincides with the conformal one, and the natural choice is the minimal one $\xi=0$. In fact, a different choice in $d=2$ would change the density matrix in an important way, inserting an additional operator $e^{\xi \phi^2}$ at the boundary. This probably will incorrectly change the value of the entropy for an interval \cite{otro1}. 

Hence, we expect the canonical stress tensor in the expression of the half boost generator (\ref{case}) to give the correct modular Hamiltonian for the free scalar field. When we transform Rindler modular Hamiltonian to a sphere of radius $R$ in the massless case,  the result will look like containing a boundary term. To make such conformal transformation we can write the Rindler modular Hamiltonian as
\be
K_0=K_{\xi^c}+2 \pi \xi^c \int_{x^1=0}d^{d-2}x\,\phi^2(x)\,,
\ee
and conformally transform it to the sphere. $K_{\xi^c}$ is transformed into a generator of conformal transformations which keep the sphere fixed and the boundary term maps to a boundary term given by an integral on the surface $\Sigma$ of the sphere,
\be
K_{\textrm{sphere}}=2\pi\int_{r<R}d^{d-1}x\, \frac{R^2-r^2}{2R} T_{00}(x,\xi^c)+2\pi \xi^c \int_{\Sigma} d\sigma \,\phi^2(x)\,.\label{hiop}
\ee
This is the result found by Herzog doing numerical calculations of $\Delta S$ for a massless scalar between a thermal state at low temperature and the vacuum \cite{herzog}. The equation $\Delta S=\Delta K$ holds for $\Delta K$ with the form (\ref{hiop}) in this example. This confirms the minimally coupled choice is the right one for Rindler space. This is also consistent with an argument by Lewkowycz and Perlmutter \cite{modu2} involving matching correlators of the modular Hamiltonian with the behavior of the Renyi entropies $S_n$ for $n$ near $1$.   

However, for an interacting conformal theory with unique stress tensor, the modular Hamiltonian must be constructed necessarily with the conformal stress tensor. Note that to construct a generic boundary term formed by an integral of a local operator $\int d^{d-2}x\,\Phi(x)$ the theory must contain an uncharged operator of dimension exactly $\Delta=d-2$. This is generically not the case except for free fields. Hence we do not expect boundary terms in the interacting case, nor for a Dirac field. A similar dichotomic story between free and interacting fixed points was recently found for the modular Hamiltonians of null slabs \cite{bousso}. For non conformal theories the presence of masses should not produce additional boundary terms because these would change the modular Hamiltonian at the $x^1=0$ corner where the density matrix is in principle dominated by the physics at the UV. 

\subsection{Free fermion}
Let us consider a free fermion field. This is the simplest case, since the stress tensor is unique. 
We have for the trace
 \be
 \Theta(x)=m \bar{\Psi}(x)\Psi(x)\,.
 \ee
 This vanishes for $m=0$, corresponding to the conformal stress tensor. 
 Using Wick's theorem we get
\bea
\mu&=&-\frac{\pi}{d(d-1)(d-2)}\int d^{d}x\,\,x^2 \,\langle \Theta(0)\Theta(x)\rangle\nonumber\\
&=&\frac{\pi m^2}{d(d-1)(d-2)}\int d^dx\,x^2\, \frac{d^dp}{(2\pi)^d}\, \frac{d^dq}{(2\pi)^d}\, \textrm{tr}\left(\frac{i\slash\!\!\! p-m}{p^2+m^2}\frac{i\slash\!\!\!q-m}{q^2+m^2}\right)  e^{i p x} e^{i q x}\,.
\eea
Now we replace $x^2$ in the integrand by $-i\nabla_p \cdot i\nabla_q$ applied to the phase factor, and then integrating by parts apply this differential operator to the rational function of $p$ and $q$. After that we are free to perform the integration in $x$ which gives a $\delta(p+q)$, and this eliminates one of the momentum integrals
\bea
\mu&=&\frac{\pi m^2}{d(d-1)(d-2)}\int d^dx\, \frac{d^dp}{(2\pi)^d}\, \frac{d^dq}{(2\pi)^d}\, \textrm{tr}\left(\frac{i\slash\!\!\! p-m}{p^2+m^2}\frac{i\slash\!\!\!q-m}{q^2+m^2}\right)(-i\nabla_p e^{i p x})(i\nabla_q e^{i q x})\nonumber\\
&=&\frac{\pi \,d_{\Psi} m^2}{d(d-1)(d-2)}\int d^dx\, \frac{d^dp}{(2\pi)^d}\, \frac{d^dq}{(2\pi)^d}\,e^{-i(p-q)x} \nabla_p\nabla_q\left(\frac{-p\cdot q+m^2}{(p^2+m^2)(q^2+m^2)}\right)\nonumber\\
&=&\frac{d_{\Psi}}{d(d-1)(d-2)2^{d-1}\pi^{(d/2-1)}\Gamma[d/2]}\int_0^\infty dp\,p^{d-1}\left(\frac{8 m^4 p^2}{(p^2+m^2)^4}-\frac{d m^2}{(p^2+m^2)^2}\right)\,,\label{fermion}
\eea
where we have introduced the dimension of the spinor space $d_\Psi$. 

In order to cutoff the integrals we can introduce for example a small distance cutoff in the integral on the first line in (\ref{fermion}), integrating for $|x|>\epsilon$, or introduce a large momentum cutoff in the last line of (\ref{fermion}), integrating for $|p|<\Lambda$, or just use dimensional regularization in this last expression in momentum space. All these calculations give the same result for the universal terms (and of course different results for the non universal ones). The result is given more compactly using dimensional regularization, that leads to 
\be
\mu_F=\frac{d_\Psi}{2} \frac{m^{d-2} \Gamma[1-d/2]}{3  \pi^{d/2-1}2^d}
\ee
for any dimension $d$. Expanding around specific dimensions we get as expected 
\begin{equation}
\mu_F=\frac{d_\Psi}{2} \mu_S
\end{equation}
with $\mu_S$ given by (\ref{evend}) and (\ref{oddd}) for even and odd dimensions, respectively. Then, this coincides both with the result of the heat kernel method in \cite{powers2}, and with the calculation of mutual information in section 2. An evaluation of the RS formula for the free fermion through the integral of the spectral function was done in \cite{rs1} with the same result.

As mentioned in section 4, the integral (\ref{fermion}) is convergent in $d=3$, and gives a negative result $\mu_F=-\frac{d_\Psi}{2} \frac{m}{12}$.

\subsection{Free scalar field}

The modular Hamiltonian and the operator $\Theta(x)$ in (\ref{47}) both involve the canonical stress tensor. Hence, the derivation of AZ formula from RS  proceeds as in section 3. The calculation of the universal area term is essentially the one in \cite{rs1}, but here we perform it using  correlators of $\Theta(x)$. We have also explained why the calculation involves the minimally coupled stress tensor, and that this is not one possible choice, but the only one giving a sensible result for the free scalar.  

We have
\be
\Theta(x)=-\left(\frac{d}{2}-1\right) (\nabla \phi)^2-\frac{d}{2} m^2 \phi^2(x)\,.
\label{traza1}
\ee
Using Wick's theorem it follows that
\bea
&-&\frac{\pi}{d(d-1)(d-2)}\int d^{d}x\,\,x^2 \,\langle \Theta(0)\Theta(x)\rangle=\nonumber\\
 &&-\frac{\pi}{d(d-1)(d-2)}\int d^dx\,x^2 \frac{d^dp}{(2\pi)^d}\, \frac{d^dq}{(2\pi)^d}\,e^{i (p+q) x} \frac{2\left((\frac{d}{2}-1)p\cdot q-\frac{d}{2}m^2\right)^2}{(p^2+m^2)(q^2+m^2)}\,.\label{88}
\eea
As for the fermion case, we replace $x^2$ in the integrand by $-\nabla_p\cdot\nabla_q$ and integrate by parts. At the end we get an integral over only one momentum,
 \bea
 \mu=&-&\frac{\pi}{d(d-1)(d-2)}\left(\frac{2 \pi^{d/2}}{\Gamma[d/2]}\right)(2\pi)^{-d}  \int_0^\infty dp\, p^{d-1}\left(\frac{-4 m^4 p^2+8 m^2 p^4+4 p^6}{(p^2+m^2)^4}\right.
\nonumber \\
 &&\left. \, +\frac{8 d p^2}{(p^2+m^2)^3}+\frac{d^2( 2 m^2+5 p^2)}{(p^2+m^2)^2}-\frac{d^3
 }{(m^2+p^2)}\right)\,.
 \eea
The integral can be performed in $d\in(0,2)$ and then continued analytically in $d$ giving
\be
\mu_S=\frac{m^{d-2} \Gamma[1-d/2]}{3  \pi^{d/2-1}2^d}\,.\label{muscalar}
\ee
Expanding around the different dimensions this gives the right numbers (\ref{evend}) and (\ref{oddd}) for even and odd dimensions respectively. The result for the universal terms are the same if we use a distance or momentum cutoff. For $d=3$ we have $\mu_S=-\frac{m}{12}$, which is negative as expected. However, the integral (\ref{88}) is not convergent, due to the fact that the stress tensor is not the conformally coupled one, and $\Theta(x)$ in (\ref{traza1}) contains a term that does not vanish for $m=0$.

\section{Renormalization of the Newton constant}
Let us review the calculation of the renormalization of Newton constant for a scalar \cite{hirohumi} to highlight the differences with the area term in EE. For the fermion field there are no couplings with the curvature and these two quantities simply coincide. This discussion will also serve to understand the role of contact terms for non-minimally coupled stress tensors which will be useful in the next section where we discuss the area term in EE for some interacting models.  

The renormalization of $(4G)^{-1}$  is given by (\ref{AZ}). Now we allow the stress tensor to have an arbitrary coupling $\xi$. Since this is related to the way the theory couples to gravity, with action (signature $(- +  + \hdots)$)
\be
{\cal A}=-\int d^dx\,\sqrt{-g}\,\frac{1}{2}\left(\partial_\mu \phi \partial^\mu \phi+m^2 \phi^2 + \xi R \phi^2\right)\,,
\ee 
the parameter $\xi$ will appear in the expression for the renormalization of $(4G)^{-1}$. 
From this action we get   
\be
\Theta(x)=-\left(\frac{d}{2}-1-2 (d-1)\xi\right) (\nabla \phi)^2-\frac{d}{2} m^2 \phi^2(x)+2(d-1)\xi \phi\nabla^2 \phi-\frac{(d-2) \xi }{2} \phi^2 R\,,
\label{traza}
\ee
where the stress tensor in the AZ formula is given by its curved space definition
\be
T_{\mu\nu}=-\frac{2}{\sqrt{-g}}\frac{\delta {\cal A}}{\delta g^{\mu\nu}}\,.
\ee

The calculation of the first term in (\ref{AZ}) proceeds as in the previous section, with $\Theta$ given by (\ref{traza}) in the flat space limit with $R=0$.  In the third term of (\ref{traza}) we can be tempted to use the equation of motion $(-\nabla^2+m^2)\phi=0$. However, in the computation of (\ref{AZ}) the   correlation of $\langle \nabla^2\phi(0)\nabla^2\phi(x)\rangle$ contains a second derivative of the delta function. Since the integrand contains a factor $x^2$, this contact term will give an additional contribution. If we evaluate $\Theta(x)$ on the equations of motion the contact term disappear and we will get an incorrect result for the renormalization of Newton's constant. We will come back to this calculation with the operator evaluated on the equations of motion in the next subsection. Note that the contact term issue did not arise for the minimally coupled scalar calculation of the previous section since $\Theta$ does not contain the term proportional to $\phi(x) \nabla^2\phi(x)$ in this case.   

We get 
 \bea
&-&\frac{\pi}{d(d-1)(d-2)}\int d^{d}x\,\,x^2 \,\langle \Theta(0)\Theta(x)\rangle=\nonumber\\
 &-&\frac{\pi}{d(d-1)(d-2)}\left(\frac{2 \pi^{d/2}}{\Gamma[d/2]}\right)(2\pi)^{-d}  \int_0^\infty dp\, p^{d-1}\left(\frac{-4 m^4 p^2+8 m^2 p^4+4 p^6}{(p^2+m^2)^4}\right.\nonumber
 \\
 &&\left. \, +\frac{8 d(1-\xi) p^2}{(p^2+m^2)^3}-\frac{d^2((-2+4 \xi)m^2+(12 \xi-5)p^2)}{(p^2+m^2)^2}+\frac{d^3(4 \xi-1)
 }{(m^2+p^2)}\right)\,.
 \eea
Even if this integral apparently depends on $\xi$, the universal part evaluated with dimensional regularization is independent of $\xi$, and gives the same result $\mu_S$ as for the minimally coupled case, eq. (\ref{muscalar}). 

This result depends crucially on the contact term and we do not get the same result if we impose a distance cutoff integrating $x^2 \langle \Theta(x)\Theta(0) \rangle$ for $|x|>\epsilon$.

For the other term in AZ formula (\ref{AZ}) we have for the operator ${\cal O}$
\begin{equation}
{\cal O}=\frac{\partial \Theta}{\partial R}=-\frac{(d-2) \xi }{2} \phi^2\,.
\end{equation}
Then the contribution is\footnote{This term has been connected to a term in Wald's entropy of black holes by several authors, see for example \cite{fursaev,malda1}.}
\begin{equation}
\frac{4 \pi}{d-2} \langle {\cal O}\rangle=-2\pi\xi\langle\phi^2\rangle=-2\pi\xi\int \frac{d^d p}{(2\pi)^d}\, \frac{1}{p^2+m^2}=(-6 \xi)\frac{m^{d-2} \Gamma[1-d/2]}{3  \pi^{d/2-1}2^d}\,.\label{po}
\end{equation}
Hence the full correction of the Newton constant is given by
\be
\Delta(4G)^{-1}=(1-6 \xi)\frac{m^{d-2} \Gamma[1-d/2]}{3  \pi^{d/2-1}2^d}=(1-6 \xi)\mu_S\,.\label{rety}
\ee
As expected this depends on $\xi$, in contrast to the universal part of the QFT entanglement entropy. This result coincides with the usual calculation of renormalization of $(4G)^{-1}$ using the heat kernel method \cite{birrell}. The result (\ref{rety}) can be unphysical if is interpreted as area term in EE, for example, it gives the wrong sign for the area term renormalization in $d=3$ for $\xi>1/6$. 
  
\subsection{Setting $\Theta$ on the equations of motion}
Now, following \cite{hirohumi} we analyze what happens if we apply the equation of motion for the field 
\be
(-\nabla^2+m^2+\xi R)\phi=0 
\ee
in the expression for $\Theta$. This has two consequences. On the one hand we replace $\nabla^2\phi$ by $m^2 \phi+\xi R \phi$ in the equation  
(\ref{traza}). This will eliminate the effect of the contact term in the term involving correlators of $\Theta$. On the other hand, this at the same time changes the dependence of $\Theta$ on $R$ and leads to a change in the term  (\ref{po}),
\be
\frac{4\pi}{d-2}\frac{\partial \Theta}{\partial R}=4 \pi\frac{2 (d-1) \xi^2}{d-2} \langle\phi^2\rangle-2\pi \xi\langle\phi^2\rangle =4\pi\frac{\xi^2}{2 \xi_c}\langle\phi^2\rangle-2\pi \xi\langle\phi^2\rangle
\ee
with $\xi_c$ the conformally coupled value (\ref{confor}). It turns out that Newton constant renormalization does not change under this transformation. That is,  
\begin{equation}
-\frac{\pi}{d(d-1)(d-2)}\int d^{d}x\,\,x^2 \,\langle 0| \hat{\Theta}(0)\hat{\Theta}(x)| 0\rangle+4\pi\frac{\xi^2}{2 \xi_c}\langle\phi^2\rangle-2\pi \xi\langle\phi^2\rangle\,,\label{tyy}
\end{equation}
with $\hat{\Theta}$ on the equations of motion, gives the correct result for all $\xi$. 

To check this we get from the previous calculation (\ref{po})
\be
4\pi\frac{\xi^2}{2 \xi_c}\langle\phi^2\rangle=6\frac{\xi^2}{\xi_c}\mu_S\,.
\ee
The first term of (\ref{tyy}) with
\be
\hat{\Theta}(x)=-\left(d/2-1-2 (d-1)\xi\right) (\nabla \phi)^2+(2(d-1)\xi-d/2) m^2 \phi^2\,,
\label{traza2}
\ee
 gives
\be
-\frac{\pi}{d(d-1)(d-2)}\int d^{d}x\,\,x^2 \,\langle 0| \hat{\Theta}(0)\hat{\Theta}(x)| 0\rangle=\mu_S-6\frac{\xi^2}{\xi_c} \mu_S\,.\label{pio}
\ee
Combining the two terms, the full  $\Delta (4 G)^{-1} =(1-6 \xi)\mu_S$ is equal to the one computed above.

This shows that AZ formula is robust under interpretations of the contact terms provided we adequately take into account the role of the second term in (\ref{AZ}). Remarkably, the correction of Newton's constant for a conformally coupled field $\xi=\xi_c$ is the same as the term depending on $\langle\Theta(0) \Theta(x)\rangle$ without taking into account contact terms, or what is the same, with the operators evaluated on the equations of motion (compare (\ref{rety}) and (\ref{pio})).

\section{Interacting examples}

\subsection{$O(N)$ model}
The universal correction for the area term of a $O(N)$ model at the Wilson Fisher fixed point at the UV was computed in \cite{on} at the lowest order in the $\epsilon=4-d$ expansion for $d=3$ (this $\epsilon$ should not be confused with the distance cutoff we used in other sections of the paper). The model has Euclidean Lagrangian
\be
{\mathcal L}=\frac{1}{2} (\partial \phi)^2+\frac{t}{2} \phi^2+\frac{u}{4} \phi^4\,,
\ee
for an $N$ dimensional vector of scalars $\phi$. For the Gaussian fixed point they find
\be
\mu=-\frac{N}{24 \pi \epsilon} m\,.\label{gaussian}
\ee
This follows from the free scalar formula (\ref{muscalar}) by replacing $d=4-\epsilon$ and expanding the coefficient of the mass around dimension $4$   \cite{on}. Hence, this is the first approximation to the correct result for $d=3$ ($\epsilon=1$), 
\be
\mu=-\frac{N}{12} m\,.
\ee
Note that the $\epsilon^{-1}$ behavior around $\epsilon=0$ corresponds to the expansion around $d=4$ where the gamma function in (\ref{muscalar}) has a divergence giving place to the universal logarithmic term in this dimension.  

For the interacting Wilson Fisher fixed point they find that
\be
\mu=-\frac{N}{144 \pi}m\,,\label{ini}
\ee
to leading order in $\epsilon$, where $m$ is the gap of the excitations. Note the result is parametrically suppressed in $\epsilon$ with respect to the Gaussian case (\ref{gaussian}). Interactions are perturbative in $\epsilon$ for the interacting fixed point, for example, the scalar dimension is
$\Delta=1-(3 \epsilon)(N+8)$ for the interacting case as opposed to $1-\epsilon/2$ for the Gaussian model. One could have naively expected the result for $\mu$ would have coincided with the Gaussian one (\ref{gaussian}) to leading order,  and differences to show up for the subleading terms in $\epsilon$. The authors arrive to this result by the replica methods for a half space and a careful treatment of the partition function in presence of a conical singularity. They find that,  in contrast to the Gaussian case, in the interacting case the partition function develops an effective operator $\sim\int d^{d-2}x \, \phi^2$ localized at the singular surface, and this insertion is the one responsible for the different parametric dependence on $\epsilon$.  

Here we can obtain this result directly from the formula (\ref{fgf}) for $\mu$ as follows. We have argued that for a theory with an interacting UV fixed point, the modular Hamiltonian for Rindler space is simply given in terms of the unique stress tensor of the theory. 
The area term is given by the formula (\ref{fgf}) involving the correlator of $\Theta(x)$.  
The stress tensor is conformal for $m=0$ in the interacting case. To lowest order in the $\epsilon$ expansion, which controls the interactions, this must be the stress tensor for free scalars but improved such that is conformal in the massless limit. Hence, to obtain the lowest order in $\epsilon$ for the area term we have to use (\ref{fgf}) where $\Theta(x)$ is the trace of the stress tensor for a scalar with $\xi=\xi_c$. We have already done this calculation in our discussion of the renormalization of the Newton constant in section 6. For $\xi=\xi_c$ formula (\ref{pio}) gives 
\be
\mu=(1-6 \xi_c)\, \mu_S=-\frac{(d-4)\pi^{1-d/2}\Gamma[1-d/2]}{3(d-1)2^{d+1}} m^{d-2}\,.\label{puo}
\ee  
Including a factor of $N$ to account for the $N$ scalars, setting $d=4-\epsilon$,  and expanding to lowest order in $\epsilon$ we get precisely (\ref{ini}). The additional factor $(1-6 \xi_c)$ in (\ref{puo}) with respect to the free scalar vanishes for $d=4$, eliminating the power $\epsilon^{-1}$ of the free case. It is important to emphasize that (\ref{pio}) has been computed using the equations of motion on the stress tensor, or equivalently, eliminating possible contact terms by setting a short distance cutoff on the integral over $x$ of the correlator. Hence,  we have to interpret formula (\ref{fgf}) with this distance cutoff in place,  that prevents ambiguous contact terms to show up in the universal term.

This example gives a non trivial confirmation for the formula of area terms in EE, the fact that contact terms should not be taken into account, and the interpretation of the stress tensor based on the modular Hamiltonian.

\subsection{A commentary on holographic calculations}

In \cite{powers} the authors computed area term corrections to holographic EE induced by perturbing the theory with relevant operators. These relevant operators are described in the bulk by scalar fields whose backreaction changes the geometry and the area of the minimal surfaces giving the EE \cite{rt}. They discover the puzzling result that perturbing with a dimension $\Delta=d-2$ operator, corresponding to the scalar mass term, a logarithmic correction to the area term appears for even dimensions but only for $d\ge 6$. This is in contrast to the free scalar field where logarithmic divergences also appear for $d=4$. The puzzle is that this result suggests there is a logarithmic term for the weak coupling limit of a theory with scalar masses in $d=4$ while this logarithmic term should be absent in the large coupling limit. 

However, this puzzle disappear if we apply the same reasoning that explains the result for the $O(N)$ model. The holographic models describe interacting CFT at the UV fixed point. In the weak coupling limit we should then use the conformally coupled stress tensor for the modular Hamiltonian and $\Theta(x)$. This gives the coefficient of the area term (\ref{puo}), which has an additional factor of $(d-4)$ with respect to the free scalar case. Hence, there is no logarithmic term for these theories even at small coupling, and the holographic result is consistent with this fact.       

In \cite{powers2} the authors made an explicit holographic computation of the area term for a ${\cal N}=2^*$ gauge theory in $d=4$ which is a massive deformation of ${\cal N}=4$ SYM at large $N$. This deformation gives masses to scalars and fermions in the theory and they obtain holographically
\be
\mu=\frac{N^2}{12\pi} m^2 \log(m \epsilon)\,,\label{ggyt}
\ee 
where $m$ is a mass parameter. At weak coupling the theory contains $N^2$ Dirac fermions of mass $m$ and $6 N^2$ real scalars of the same mass. As we have seen above, for models with an interacting UV fixed point, the scalars do not contribute to the logarithmic term in $d=4$ at weak coupling, and we have from (\ref{ferm}) that, at small coupling,
\be
\mu=N^2 \times \frac{1}{12\pi} m^2 \log(m \epsilon)\,,
\ee
which coincides with (\ref{ggyt}).\footnote{It was noted in \cite{powers2} that the strong coupling result coincided with the weak coupling one for the fermions alone.} The puzzle here is then why the weak coupling result coincides with the strong coupling one.

\section{Conclusions}
We have studied area terms in EE in flat space quantum field theory and shown there are universal terms that can be well defined through mutual information. These terms should be non ambiguous and cannot depend on how the theory is extended to curved space. We have argued that the formula
\be
\mu=-\frac{\pi}{d(d-1)(d-2)}\int_{|x|>\epsilon} d^{d}x\,\,x^2 \,\langle 0| \Theta(0)\Theta(x)| 0\rangle \label{opls}
\ee
gives the area terms in EE for a general QFT. For free scalar fields there are some subtleties: The minimally coupled stress tensor have to be used in (\ref{opls}). The specific distance cutoff in (\ref{opls}) is important to eliminate possible contributions from contact terms. 

This is equivalent to RS formula (\ref{RS}) provided we use the correct modular Hamiltonian and operator $\Theta$. For theories with interacting UV fixed point we expect a unique stress tensor enters into all these formulas. 

The differences between free and interacting UV fixed points imply a discontinuous change of modular Hamiltonian for some models between zero and non zero coupling. This phenomenon has also been found in terms of the replica partition function in \cite{on}.   

We have argued that an interesting consequence of (\ref{opls}) is that the constant term in the entropy of a circle in $d=3$, which acts as the monotonous quantity in the $F$ theorem, must necesarily change from the UV to the IR. This must be the case when (\ref{opls}) in $d=3$ is convergent. However, holographic calculations suggest this should be also the case for divengent $\mu$. From (\ref{opls}) we have also recovered the well known Zamolodchikov's formula for the running of the Virasoro central charge in $d=2$ QFT.  

The formula (\ref{opls}) does not give in general the renormalization of the Newton constant, eq. (\ref{AZ}), because in this later there is another term depending on the curvature couplings. Moreover the renormalization of Newton's constant depends on the contact terms in the correlators, while (\ref{opls}) does not. However, it seems that for any theory it is possible to find a specific extension of the theory to curved space where the universal correction in $(4G)^{-1}$ coincides with the one in the area terms in EE. 

It would be nice to study (\ref{opls}) holographically. It would also be interesting to try to understand the difference between area terms in EE and the AZ formula for the Newton constant thinking on the entanglement interpretation to BH entropy.\footnote{In this context see for example the recent work \cite{solo14}.} This difference may contain some clues on how gravity renormalizes EE and at the same time conserves its entropic origin.

\section*{Acknowledgements}

It is a pleasure to thank discussions with 
 Christopher Herzog, Marina Huerta, Juan Maldacena, Aitor Lewkowycz, Rob Myers, Vladimir Rosenhaus, Misha Smolkin, and Gonzalo Torroba. This work was supported by CONICET, CNEA, Universidad Nacional de Cuyo,  and ANPCyT, Argentina.

\end{document}